\documentclass[prd,aps,nofootinbib,onecolumn,showpacs,10pt]{revtex4-1}
\usepackage{graphicx,epsfig}

\usepackage{epsf}
\usepackage{graphicx}

\begin{document}

\title{  \bf   $g_{D^{\ast}_{s}D K^{\ast}(892)}$ and
$g_{B^{\ast}_{s}B K^{\ast}(892)}$ coupling constants in QCD sum
rules}
\author{K. Azizi $^{\dag1}$ ,H. Sundu $^{*2}$ \\
 $^{\dag}$Physics Division,  Faculty of Arts and Sciences,
Do\u gu\c s University,
 Ac{\i}badem-Kad{\i}k\"oy, \\ 34722 Istanbul, Turkey\\
 $^{*}$Department of Physics , Kocaeli University, 41380 Izmit,
Turkey\\
$^1$e-mail:kazizi@dogus.edu.tr\\
$^2$email:hayriye.sundu@kocaeli.edu.tr}

\begin{abstract}
The coupling constants $g_{D^{\ast}_{s}D K^{\ast}(892)}$ and
$g_{B^{\ast}_{s}B K^{\ast}(892)}$ are calculated in the framework of
three-point QCD sum rules.  The correlation functions responsible
for these coupling constants are evaluated considering
contributions of both $D(B)$ and $K^*(892)$ mesons as off-shell
states, but in the absence of radiative corrections. The results, $g_{D^{\ast}_{s}D
K^{\ast}(892)}=(3.74\pm1.38)~GeV^{-1}$ and $g_{B^{\ast}_{s}B
K^{\ast}(892)}=(3.24\pm1.08)~GeV^{-1}$ are obtained for the
considered strong coupling constants.
\end{abstract}

\maketitle

\section{Introduction}
The heavy-heavy-light mesons coupling constants are fundamental objects as they
can provide essential information on the low energy QCD. Their numerical values obtained in QCD can bring important constraints in constructing the meson-meson potentials and strong interactions among them.
 In the recent years, both theoretical and experimental studies on heavy
mesons  have received   considerable attention. In this connection, excited experimental results obtained in BABAR, FERMILAB, CLEO, CDF,
D0, etc. \cite{Babar,FermiLab,CLEO,CDF,CDF1,D0,CDFD0,CERN,CDF2}
 and some physical properties  of these mesons have been
studied using various theoretical models (see for instance
\cite{CTHDavies,GLi,DMelikhov,AMBadalian,Kazim,EBGregory}).

In this article, we calculate the strong coupling constants,   $g_{D^{\ast}_{s}D K^{\ast}(892)}$ and
$g_{B^{\ast}_{s}B K^{\ast}(892)}$ in the framework of three-point QCD sum rules considering
contributions of both $D(B)$ and $K^*(892)$ mesons as off-shell
states, but in the absence of radiative corrections.  The result of these coupling constants can help us to better analyze the results of existing experiments hold at different centers.
For instance, consider the $B_c$ meson or
the newly discovered charmonium states, $X$, $Y$ and  $Z$  by   BABAR and BELLE collaborations. These states  decay to a $J/\psi$ or $\psi'$ and a light meson in the final state. However, it is supposed that
first
 these states decay into an intermediate two body states containing $D_q$ or  $D^*_q$ with $q=u,~d$ or $s$ quarks, then these intermediate states decay into
final stats with the exchange of one or more virtual mesons. The similar procedure may happen in decays of heavy bottonium. Hence, to get precise information about such transitions, we need to have information about the coupling constants between participating particles.

Calculation of the heavy-heavy-light mesons coupling constants via the fundamental theory of
QCD is highly valuable. However, such interactions lie in a region very far from the
perturbative regime, hence the fundamental QCD Lagrangian can not be responsible in this respect. Therefore, we need some non-perturbative approaches like QCD
sum rules \cite{Shifman} as one of the most powerful and applicable tools to hadron physics. Note that, the coupling constants,
$D^*D_sK$, $D^*_sDK$ \cite{MEBracco,ZGWang}, $D_0D_sK$,  $D_{s0}DK$
\cite{ZGWang}, $D^*D\rho$ \cite{BORodrigues}, $D^*D\pi$
\cite{FSNavarra, FSNavarra1}, $DD\rho$ \cite{MEBracco1}, $DDJ/\psi$
\cite{RDMatheus}, $D^*DJ/\psi$ \cite{RDMatheus1}, $D^*D^*\pi$
\cite{ZGWang1,FCarvalho}, $D^*D^*J/\psi$ \cite{MEBracco2},
$D_sD^*K$, $D_s^*DK$ \cite{MEBracco3}, $DD\omega$ \cite{LBHolanda},
$D^*D^*\rho$ \cite{MEBracco4}, and $B_{s0}BK$,
$B_{s1}B^*K$ \cite{ZGWang2} have been investigated using different approaches.

 This paper is organized as
follows. In section 2, we give the details of QCD sum rules
 for the considered coupling constants when both $D(B)$ and $K^*(892)$ mesons in the final state are off-shell. The next section is devoted to  the
numerical analysis and discussion.

\section{QCD Sum Rules for the Coupling Constants}

In this section, we derive  QCD sum rules for coupling constants. For
this aim, we will evaluate the three-point correlation functions,
\begin{eqnarray}\label{CorrelationFunc1}
\Pi_{\mu\nu}^{D (B)}(p^{\prime},q)=i^2 \int d^4x~d^4y~
e^{ip^{\prime}\cdot x}~ e^{iq\cdot y}{\langle}0| {\cal T}\left (
\eta^{K^{\ast}}_{\nu}(x)~ \eta^{D(B)}(y)~
\eta^{D_{s}^{\ast}(B_{s}^{\ast})\dag}_{\mu}(0)\right)|0{\rangle}
\end{eqnarray}
for  $D(B)$ off-shell, and
\begin{eqnarray}\label{CorrelationFunc2}
\Pi_{\mu\nu}^{K^{\ast}(892)}(p^{\prime},q)=i^2 \int d^4x~d^4y~
e^{ip^{\prime}\cdot x}~ e^{iq\cdot y}{\langle}0| {\cal T}\left (
\eta^{D(B)}(x)~ \eta^{K^{\ast}}_{\nu}(y)~
\eta^{D_{s}^{\ast}(B_{s}^{\ast})\dag}_{\mu}(0)\right)|0{\rangle}
\end{eqnarray}
for  $K^{\ast}(892)$ off-shell. Here ${\cal T}$ is  the time
ordering operator and $q=p-p'$ is transferred momentum. Each meson interpolating field can be written in
terms of the quark field operators as following form:
\begin{eqnarray}\label{mesonintfield}
\eta^{K^{\ast}}_{\nu}(x)&=& \overline{s}(x)\gamma_{\nu}u(x)
\nonumber \\
\eta^{D[B]}(y)&=& \overline{u}(y)\gamma_{5}c[b](y)
\nonumber \\
\eta^{D_{s}^{\ast}[B_{s}^{\ast}]}_{\mu}(0)&=&
\overline{s}(0)\gamma_{\mu}c[b](0)
\end{eqnarray}
The correlation functions are calculated in two different ways. In phenomenological or physical side, they are obtained in terms of hadronic parameters. In theoretical or  QCD  side,
they are evaluated in terms of quark and gluon degrees of freedom by the help
of the operator product expansion (OPE) in deep Euclidean region. The sum rules for the coupling constants are obtained equating the coefficient of a sufficient structure from both sides of the same correlation functions.
 To suppress  contribution of the higher states and continuum,  double Borel transformation with respect to the variables, $p^2$ and $p'^2$ is applied.

First, let us focus on the calculation of the physical side of the first
correlation function (Eq.(\ref{CorrelationFunc1})) for an off-shell
$D(B)$ meson. The physical part is obtained by saturating
Eq. (\ref{CorrelationFunc1}) with the complete sets of appropriate $D^0$, $D_s^{\ast}$
and $K^{\ast}(892)$ states with the same quantum numbers as the corresponding interpolating currents. After performing four-integrals over $x$ and $y$,
we obtain:
\begin{eqnarray}\label{CorrelationFuncPhys1}
\Pi_{\mu\nu}^{D
(B)}(p^{\prime},q)&=&\frac{{\langle}0|\eta^{K^{\ast}}_{\nu}|K^{\ast}
(p^{\prime},\epsilon){\rangle} {\langle}0|\eta^{D(B)}|D(B)
(q){\rangle} {\langle}K^{\ast}(p^{\prime},\epsilon) D(B)
(q)|D_{s}^{\ast}(B_{s}^{\ast})(p,\epsilon^{\prime}){\rangle}
{\langle}D_s^{\ast}(B_{s}^{\ast})
(p,\epsilon^{\prime})|\eta^{D_s^{\ast}(B_{s}^{\ast})}_{\mu}|0{\rangle}}
{(q^2-m_{D(B)}^2)
(p^2-m_{D_s^{\ast}(B_{s}^{\ast})}^2)({p^{\prime}}^{2}-m_{K^{\ast}}^2)}
\nonumber \\ &+&...
\end{eqnarray}
where .... represents the contribution of the higher states and
continuum. The  matrix elements appearing in the  above
equation are defined in terms of hadronic parameters such as masses, leptonic
decay constants and coupling constant, i.e.,
\begin{eqnarray}\label{transitionamp}
{\langle}
0|\eta_{\nu}^{K^{\ast}}|K^{\ast}(q,\epsilon){\rangle}&=&m_{K^{\ast}}f_{K^{\ast}}\epsilon_{\nu}
\nonumber\\
{\langle}0|\eta^{D(B)}|D(B)(p^{\prime}){\rangle}&=&i\frac{m_{D(B)}^2~f_{D(B)}}{m_{c(b)}+m_u}
\nonumber\\
{\langle}D_{s}^{\ast}(B_{s}^{\ast})(p,\epsilon^{\prime})
 |\eta^{D_{s}^{\ast}(B_{s}^{\ast})}_{\mu}|0{\rangle}&=&m_{D_{s}^{\ast}(B_{s}^{\ast})}
 f_{D_{s}^{\ast}(B_{s}^{\ast})}{\epsilon^{\ast}}^{\prime}_{\mu}
 \nonumber\\
{\langle}K^{\ast}(q,\epsilon)D(B)(p^{\prime})|D^{\ast}_{s}(B^{\ast}_{s})(p,\epsilon^{\prime}){\rangle}&=&ig^{D(B)}_{D^{\ast}_{s}D
K^{\ast}(B^{\ast}_{s}B
K^{\ast})}\varepsilon^{\alpha\beta\eta\theta}\epsilon^{\ast}_{\theta}\epsilon^{\prime}_{\eta}p^{\prime}_{\beta}p_{\alpha}
\end{eqnarray}
where $g^{D(B)}_{D^{\ast}_{s}D
K^{\ast}(B^{\ast}_{s}B
K^{\ast})}$ is coupling constant when $D(B)$ is off-shell and $\epsilon$ and $\epsilon^{\prime}$ are the polarization
vectors associated with the $K^{\ast}$ and $D_{s}^{\ast}(B^{\ast}_{s})$,
respectively. Using
Eq. (\ref{transitionamp}) in Eq. (\ref{CorrelationFuncPhys1}) and summing over polarization vectors via,
\begin{eqnarray}\label{polvec}
\epsilon_\nu\epsilon^*_\theta=-g_{\nu\theta}+\frac{q_{\nu}q_{\theta}}{m_{K^{\ast}}^2},\nonumber\\
\epsilon'_\eta\epsilon^{'*}_\mu=-g_{\eta\mu}+\frac{p_{\eta}p_{\mu}}{m_{D_{s}^{\ast}(B_{s}^{\ast})}^2},
\end{eqnarray}
the physical
side of the correlation function for $D(B)$  off-shell  is
obtained as:
\begin{eqnarray}\label{CorrelationFuncPhys2}
\Pi_{\mu\nu}^{D (B)}(p^{\prime},p)&=&-g^{D(B)}_{D^{\ast}_{s}D
K^{\ast}(B^{\ast}_{s}B
K^{\ast})}(q^2)\frac{f_{D^{\ast}_{s}(B^{\ast}_{s})}f_{D(B)}
f_{K^{\ast}}\frac{m_{D(B)}^2}{m_{c(b)+m_u}}m_{D^{\ast}_{s}(B^{\ast}_{s})}
m_{K^{\ast}}}{(q^2-m_{D(B)}^2)
(p^2-m_{D_s^{\ast}(B_{s}^{\ast})}^2)({p^{\prime}}^{2}-m_{K^{\ast}}^2)}\varepsilon^{\alpha\beta\mu\nu}
p^{\prime}_{\beta}p_{\alpha}+ ....
\end{eqnarray}
To calculate the coupling constant, we will choose the structure, $\varepsilon^{\alpha\beta\mu\nu}
p^{\prime}_{\beta}p_{\alpha}$ from both sides of the correlation functions.
From a similar way, we obtain the final expression of the physical side of the correlation function for an off-shell $K^{\ast}$ meson  as:

\begin{eqnarray}\label{CorrelationFuncPhys22}
\Pi_{\mu\nu}^{K^{\ast}}(p^{\prime},p)&=&-g^{K^{\ast}}_{D^{\ast}_{s}D
K^{\ast}(B^{\ast}_{s}B
K^{\ast})}(q^2)\frac{f_{D^{\ast}_{s}(B^{\ast}_{s})}f_{D(B)}
f_{K^{\ast}}\frac{m_{D(B)}^2}{m_{c(b)+m_u}}m_{D^{\ast}_{s}(B^{\ast}_{s})}
m_{K^{\ast}}}{({p^{\prime}}^{2}-m_{D(B)}^2)
(p^2-m_{D_s^{\ast}(B_{s}^{\ast})}^2)(q^2-m_{K^{\ast}}^2)}\varepsilon^{\alpha\beta\mu\nu}
p^{\prime}_{\beta}p_{\alpha}+....
\end{eqnarray}
\begin{figure}[h!]
\begin{center}
\includegraphics[width=12cm]{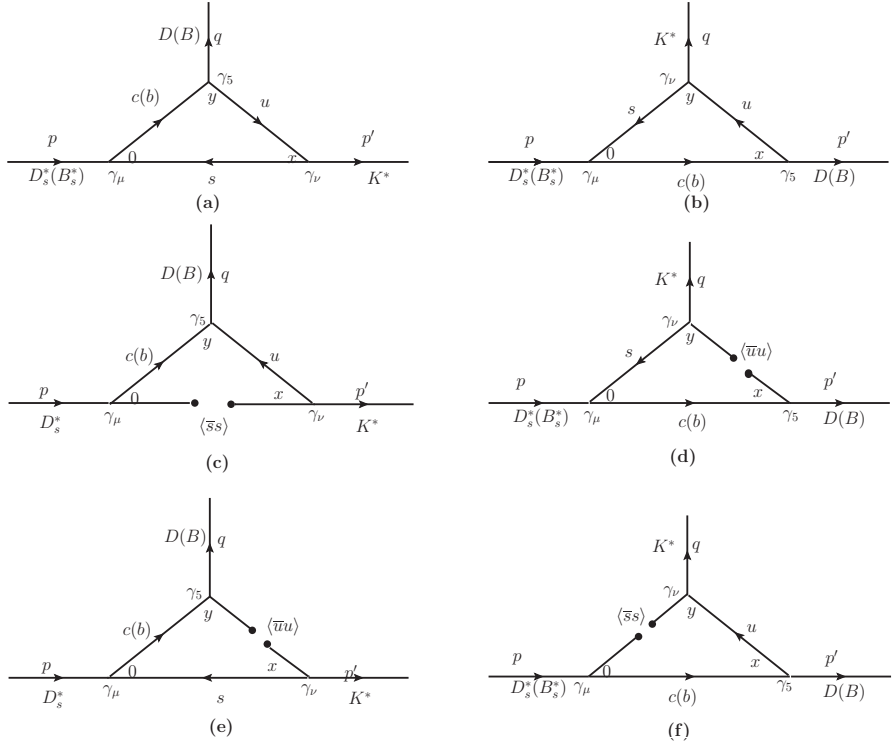}
\end{center}
\caption{(a) and (b): Bare loop diagram for the $D(B)$  and
$K^{\ast}$ off-shell, respectively; (c) and (e): Diagrams
corresponding to quark condensate for the $D(B)$ off-shell; (d) and
(f): Diagrams corresponding to quark condensate for the  $K^{\ast}$
off-shell.} \label{Figure}
\end{figure}

Now, we concentrate our attention to calculate the QCD or theoretical  side of the correlation
functions  in deep Euclidean space, where
$p^2\rightarrow-\infty$ and ${p^{\prime}}^2\rightarrow-\infty$. For this aim, each correlation function, $\Pi^{i}_{\mu\nu}(p^{\prime},p)$, where $i$
stands for $D(B)$ or
$K^{\ast}$,  can be written in terms of
perturbative and non-perturbative parts as:
\begin{eqnarray}\label{CorrelationFuncQCD}
\Pi^{i}_{\mu\nu}(p^{\prime},p)&=&
\left(\Pi_{per}+\Pi_{nonper}\right)\varepsilon^{\alpha\beta\mu\nu}
p^{\prime}_{\beta}p_{\alpha},
\end{eqnarray}
where the  perturbative part is defined in terms of double dispersion
integral as:
\begin{eqnarray}\label{CorrelationFuncQCDPert}
\Pi_{per}&=&-\frac{1}{4 \pi^{2}} \int ds^{\prime} \int ds
\frac{\rho(s,s^{\prime}, q^2)}{(s-p^2)
(s^{\prime}-{p^{\prime}}^2)}+\mbox{subtraction terms},
\end{eqnarray}
here, $\rho(s,s^{\prime}, q^2) $ is called spectral density. In
order to obtain the spectral density, we need to calculate the bare
loop diagrams (a) and (b) in Fig.(\ref{Figure}) for $D(B)$ and
$K^{\ast}$ off-shell, respectively. We calculate these
diagrams in terms of the usual Feynman integral by the help of
Cutkosky rules, i.e., by replacing the quark propagators with Dirac
delta functions: $\frac{1}{q^2-m^2}\rightarrow (- 2\pi i)
\delta(q^2-m^2)$. After some straightforward calculations, we obtain
the spectral densities as following:
\begin{eqnarray}\label{SpecDenstD}
\rho^{D}(s,s^{\prime},q^2)&=&\frac{N_c}{\lambda^{3/2}(s,s^{\prime},q^2)}
\left\{2 m_s^3 q^2+m_u s \left(2 m_u^2-q^2+s-s^{\prime}\right)-m_s^2
m_u \left(q^2+s-s^{\prime}\right)+2 m_c^3 s^{\prime}+m_c^2 \left[m_s
\left(-q^2+s-s^{\prime}\right) \right.\right.
\nonumber \\
 & -&\left. m_u \left(-q^2+s+s^{\prime}\right)\right]+m_c
 \left[m_s^2
\left(-q^2+s-s^{\prime}\right)-\left(q^2+s-s^{\prime}\right)
s^{\prime}-m_u^2 \left(-q^2+s+s^{\prime}\right)\right]
\nonumber \\
&-& \left.m_s \left(m_u^2 \left(q^2+s-s^{\prime}\right)+q^2
\left[-q^2+s+s^{\prime}\right)\right]\right\},
\end{eqnarray}
\begin{eqnarray}\label{SpecDenstDKs}
\rho_1^{K^{\ast}}(s,s^{\prime},q^2)&=&\frac{N_c}{\lambda^{3/2}
(s,s^{\prime},q^2)} \left\{2 m_c^3 q^2+m_u s \left(2
m_u^2-q^2+s-s^{\prime} \right)-m_c^2 m_u
\left(q^2+s-s^{\prime}\right)+2 m_s^3 s^{\prime}+m_s^2 \left(m_c
\left(-q^2+s-s^{\prime}\right)\right.\right.
\nonumber \\
 &-&\left.m_u
\left(-q^2+s+s^{\prime}\right)\right)+m_s \left[m_c^2
\left(-q^2+s-s^{\prime}\right)-\left(q^2+s-s^{\prime}\right)
s^{\prime}-m_u^2 \left(-q^2+s+s^{\prime}\right)\right]
\nonumber \\
 &-&\left.m_c \left[m_u^2 \left(q^2+s-s^{\prime}\right)+q^2
\left(-q^2+s+s^{\prime}\right)\right]\right\},
\end{eqnarray}
for the $D_s^{\ast}D K^{\ast}(892)$ vertex associated with the
off-shell D and $K^{\ast}(892)$ meson, respectively, and
\begin{eqnarray}\label{SpecDenstB}
\rho^{B}(s,s^{\prime},q^2)&=&\frac{N_c}{\lambda^{3/2}(s,s^{\prime},q^2)}
\left\{2 m_s^3 q^2+m_u s \left(2 m_u^2-q^2+s-s^{\prime}\right)-m_s^2
m_u \left(q^2+s-s^{\prime}\right)+2 m_b^3 s^{\prime}+m_b^2 \left[m_s
\left(-q^2+s-s^{\prime}\right) \right.\right.
\nonumber \\
 & -&\left. m_u \left(-q^2+s+s^{\prime}\right)\right]+m_b \left[m_s^2
\left(-q^2+s-s^{\prime}\right)-\left(q^2+s-s^{\prime}\right)
s^{\prime}-m_u^2 \left(-q^2+s+s^{\prime}\right)\right]
\nonumber \\
&-& \left.m_s \left(m_u^2 \left(q^2+s-s^{\prime}\right)+q^2
\left[-q^2+s+s^{\prime}\right)\right]\right\},
\end{eqnarray}
\begin{eqnarray}\label{SpecDenstDKs}
\rho_2^{K^{\ast}}(s,s^{\prime},q^2)&=&\frac{N_c}{\lambda^{3/2}(s,s^{\prime},q^2)}
\left\{2 m_b^3 q^2+m_u s \left(2 m_u^2-q^2+s-s^{\prime}\right)-m_b^2
m_u \left(q^2+s-s^{\prime}\right)+2 m_s^3
s^{\prime}+m_s^2 \left(m_b
\left(-q^2+s-s^{\prime}\right)\right.\right.
\nonumber \\
 &-&\left.m_u
\left(-q^2+s+s^{\prime}\right)\right)+m_s \left[m_b^2
\left(-q^2+s-s^{\prime}\right)-\left(q^2+s-s^{\prime}\right)
s^{\prime}-m_u^2 \left(-q^2+s+s^{\prime}\right)\right]
\nonumber \\
 &-&\left.m_b \left[m_u^2 \left(q^2+s-s^{\prime}\right)+q^2
\left(-q^2+s+s^{\prime}\right)\right]\right\},
\end{eqnarray}
for the $B_s^{\ast}B K^{\ast}(892)$ vertex associated with the
off-shell B and $K^{\ast}(892)$ meson, respectively. Here
$\lambda(a,b,c)=a^2+b^2+c^2-2ac-2bc-2ab$ and $N_c=3$ is the color
number.

To calculate the nonperturbative contributions in QCD side, we
consider  the
quark condensate diagrams presented in (c), (d), (e) and (f) parts of
Fig. (\ref{Figure}). It should be reminded that the heavy quark
condensates contributions are suppressed by inverse of the heavy
quark mass, so they can be safely neglected. Therefore, as nonperturbative part, we only encounter contributions coming  from light quark
condensates. Contributions of the diagrams (d), (e) and
(f) in Fig.(\ref{Figure}) are zero since applying double Borel
transformation with respect to the both variables $p^2$ and
${p^{\prime}}^2$ will kill them because of appearing
only one variable in the denominator in these cases. Hence, we calculate the diagram (c) in
Fig.(\ref{Figure}) for the off-shell $D(B)$ meson. As a result, we obtain:
\begin{eqnarray}\label{CorrelationFuncNonpert}
\Pi_{nonper}^{D}&=&-\frac{{\langle}\overline{s}s{\rangle}}{(p^2-m_c^2)({p^{\prime}}^2-m_u^2)},
\end{eqnarray}
for the off-shell D meson and
\begin{eqnarray}\label{CorrelationFuncNonpert}
\Pi_{nonper}^{B}&=&-\frac{{\langle}\overline{s}s{\rangle}}{(p^2-m_b^2)({p^{\prime}}^2-m_u^2)},
\end{eqnarray}
for the off-shell B meson.

Now, it is time to apply the double Borel transformations with respect
to the $p^2(p^2\rightarrow M^2)$ and ${p^{\prime}}^2\rightarrow
({p^{\prime}}^2\rightarrow {M^{\prime}}^2)$ to the physical as well
as the QCD sides and equate the coefficient of the selected structure from two representations. Finally, we get the following sum rules for the corresponding coupling
 constant form factors:

\begin{eqnarray}\label{CoupCons-gDsDKs-Doffshel}
g^{D}_{D^{\ast}_{s}DK^{\ast}}(q^2)&=&\frac{(q^2-m_D^2)}{f_{D_s^{\ast}}
f_D f_{K^{\ast}}\frac{m_D^2}{m_c+m_u}m_{D_s^{\ast}}m_{K^{\ast}}}
e^{\frac{m_{D^{\ast}}^2}{M^2}}e^{\frac{m_{K^{\ast}}^2}{{M^{\prime}}^2}}
\left[\frac{1}{4~\pi^2}\int^{s_0}_{(m_c+m_s)^2}
ds\int^{s^{\prime}_0}_{(m_s+m_u)^2} ds^{\prime}
\rho^{D}(s,s^{\prime},q^2)\right.
\nonumber \\
&& \left. \theta
[1-{(f^D(s,s^{\prime}))}^2]e^{\frac{-s}{M^2}}e^{\frac{-s^{\prime}}
{{M^{\prime}}^2}}+{\langle}\overline{s}s{\rangle}
e^{\frac{m_c^2}{M^2}}e^{\frac{m_u^2}{{M^{\prime}}^2}}\right],
\end{eqnarray}
\begin{eqnarray}\label{CoupCons-gDsDKs-Ksoffshel}
g^{K^{\ast}}_{D^{\ast}_{s}DK^{\ast}}(q^2)&=&\frac{(q^2-{m_{K^{\ast}}}^2)}
{f_{D_s^{\ast}} f_D
f_{K^{\ast}}\frac{m_D^2}{m_c+m_u}m_{D_s^{\ast}}m_{K^{\ast}}}
e^{\frac{m_{D^{\ast}}^2}{M^2}}e^{\frac{m_{D}^2}{{M^{\prime}}^2}}
\left[\frac{1}{4~\pi^2}\int^{s_0}_{(m_c+m_s)^2}
ds\int^{s^{\prime}_0}_{(m_c+m_u)^2} ds^{\prime}
\rho_1^{K^{\ast}}(s,s^{\prime},q^2) \right.
\nonumber \\
&& \left. \theta
[1-{(f_1^{K^{\ast}}(s,s^{\prime}))}^2]e^{\frac{-s}{M^2}}e^{\frac{-s^{\prime}}
{{M^{\prime}}^2}}\right],
\end{eqnarray}
for the off-shell D and $K^{\ast}(892)$ meson associated with the
$D_s^{\ast}D K^{\ast}(892)$ vertex, respectively, and
\begin{eqnarray}\label{CoupCons-gBsBKs-Boffshel}
g^{B}_{B^{\ast}_{s}BK^{\ast}}(q^2)&=&\frac{(q^2-m_B^2)}{f_{B_s^{\ast}}
f_B f_{K^{\ast}}\frac{m_B^2}{m_b+m_u}m_{B_s^{\ast}}m_{K^{\ast}}}
e^{\frac{m_{B_s^{\ast}}^2}{M^2}}e^{\frac{m_{K^{\ast}}^2}{{M^{\prime}}^2}}
\left[\frac{1}{4~\pi^2}\int^{s_0}_{(m_b+m_s)^2}
ds\int^{s_0^{\prime}}_{(m_s+m_u)^2} ds^{\prime}
\rho^{B}(s,s^{\prime},q^2) \right.
\nonumber \\
&& \left.
\theta[1-{(f^B(s,s^{\prime}))}^2]e^{\frac{-s}{M^2}}e^{\frac{-s^{\prime}}
{{M^{\prime}}^2}}+{\langle}\overline{s}s{\rangle}
e^{\frac{m_b^2}{M^2}}e^{\frac{m_u^2}{{M^{\prime}}^2}}\right],
\end{eqnarray}
\begin{eqnarray}\label{CoupCons-gBsBKs-Ksoffshel}
g^{K^{\ast}}_{B^{\ast}_{s}BK^{\ast}}(q^2)&=&\frac{(q^2-{m_{K^{\ast}}}^2)}
{f_{B_s^{\ast}} f_B
f_{K^{\ast}}\frac{m_B^2}{m_b+m_u}m_{B_s^{\ast}}m_{K^{\ast}}}
e^{\frac{m_{B_s^{\ast}}^2}{M^2}}e^{\frac{m_{B}^2}{{M^{\prime}}^2}}
\left[\frac{1}{4~\pi^2}\int^{s_0}_{(m_b+m_s)^2}
ds\int^{{s_0}^{\prime}}_{(m_b+m_u)^2} ds^{\prime}
\rho_2^{K^{\ast}}(s,s^{\prime},q^2) \right.
\nonumber \\
&& \left.
\theta[1-{(f_2^{K^{\ast}}(s,s^{\prime}))}^2]e^{\frac{-s}{M^2}}e^{\frac{-s^{\prime}}
{{M^{\prime}}^2}}\right],
\end{eqnarray}
for the off-shell B and $K^{\ast}(892)$ meson associated with the
$B_s^{\ast}B K^{\ast}(892)$ vertex, respectively.
The integration regions in the perturbative part in
Eqs. (\ref{CoupCons-gDsDKs-Doffshel})-(\ref{CoupCons-gBsBKs-Ksoffshel})
are determined requiring that the arguments of the three $\delta$
functions coming from Cutkosky rule vanish simultaneously. So, the
physical regions in the $s$ - $s^{\prime}$ plane are described by the
following non-equalities:
\begin{eqnarray}\label{fsspD0offshell}
-1\leq
f^D(s,s^{\prime})=\frac{2~s~(m_s^2-m_u^2+s^{\prime})+(m_c^2-m_s^2-s)
(-q^2+s+s^{\prime})}{\lambda^{1/2}(m_c^2,m_s^2,s)
\lambda^{1/2}(s,s^{\prime},q^2)}\leq 1,
\end{eqnarray}
\begin{eqnarray}\label{fsspKs1offshell}
-1\leq
f_1^{K^{\ast}}(s,s^{\prime})=\frac{2~s~(-m_c^2+m_u^2-s^{\prime})+(m_c^2-m_s^2+s)
(-q^2+s+s^{\prime})}{\lambda^{1/2}(m_c^2,m_s^2,s)
\lambda^{1/2}(s,s^{\prime},q^2)}\leq 1,
\end{eqnarray}
for the off-shell D and $K^{\ast}(892)$ meson associated with the
$D_s^{\ast}D K^{\ast}(892)$ vertex, respectively, and
\begin{eqnarray}\label{fsspBoffshell}
-1\leq
f^B(s,s^{\prime})=\frac{2~s~(m_s^2-m_u^2+s^{\prime})+(m_b^2-m_s^2-s)
(-q^2+s+s^{\prime})}{\lambda^{1/2}(m_b^2,m_s^2,s)
\lambda^{1/2}(s,s^{\prime},q^2)}\leq 1,
\end{eqnarray}
\begin{eqnarray}\label{fsspKs2offshell}
-1\leq
f_2^{K^{\ast}}(s,s^{\prime})=\frac{2~s~(-m_b^2+m_u^2-s^{\prime})
+(m_b^2-m_s^2+s) (-q^2+s+s^{\prime})}{\lambda^{1/2}(m_b^2,m_s^2,s)
\lambda^{1/2}(s,s^{\prime},q^2)}\leq 1,
\end{eqnarray}
for the off-shell B and $K^{\ast}(892)$ meson associated with the
$B_s^{\ast}B K^{\ast}(892)$ vertex, respectively. These physical regions are imposed by the limits on the integrals and step functions in the integrands of the sum rules.
In order to subtract the contributions of the higher states and
continuum, the quark-hadron duality assumption is used, i.e., it is
assumed that,
\begin{eqnarray}\label{ope}
\rho^{higher states}(s,s') = \rho^{OPE}(s,s') \theta(s-s_0)
\theta(s'-s'_0).
\end{eqnarray}
Note that, the double Borel transformation used in calculations  is
defined as:
\begin{equation}\label{16au}
\hat{B}\frac{1}{(p^2-m^2_1)^m}\frac{1}{(p'^2-m^2_2)^n}\rightarrow(-1)^{m+n}\frac{1}{\Gamma(m)}\frac{1}{\Gamma
(n)}e^{-m_{1}^2/M^{2}}e^{-m_{2}^2/M^{'2}}\frac{1}{(M^{2})^{m-1}(M^{'2})^{n-1}}.
\end{equation}

\section{Numerical analysis}

Present section is devoted to the numerical analysis of the sum
rules for the coupling constants. In further analysis, we use,
$m_{K^{\ast}}(892)=
 (0.89166\pm0.00026)~GeV$, $m_{D^0}=(1.8648\pm0.00014)~GeV$,
 $m_{D_{s}^{\ast}}=(2.1123\pm0.0005)~GeV$, $m_{B^{\pm}}=(5.2792\pm
 0.0003)~GeV$, $m_{B_s^{\ast}}=(5.4154\pm0.0014)~GeV$ \cite{K.Nakamura},
$m_c=1.3~GeV$, $m_{b}=4.7~GeV$, $m_s=0.14~GeV$\cite{B.L.Ioffe},
$m_u=0$, $f_{K^{\ast}}=225~MeV$\cite{P.Maris},
$f_{D_s^{\ast}}=(272\pm16^{0}_{-20})~MeV$,
$f_{B_s^{\ast}}=(229\pm20^{31}_{-16})~MeV$ \cite{D.Becirevic},
$f_{B}=(190\pm13)~MeV$ \cite{E.Gamiz}, $f_{D}=(206.7\pm8.9)~MeV$
\cite{J.L.Rosner}
 and ${\langle}\overline{s}s{\rangle}=-0.8(0.24\pm0.01)^3~GeV^3$
\cite{B.L.Ioffe}.

\begin{figure}[h!]
\begin{center}
\includegraphics[width=13cm]{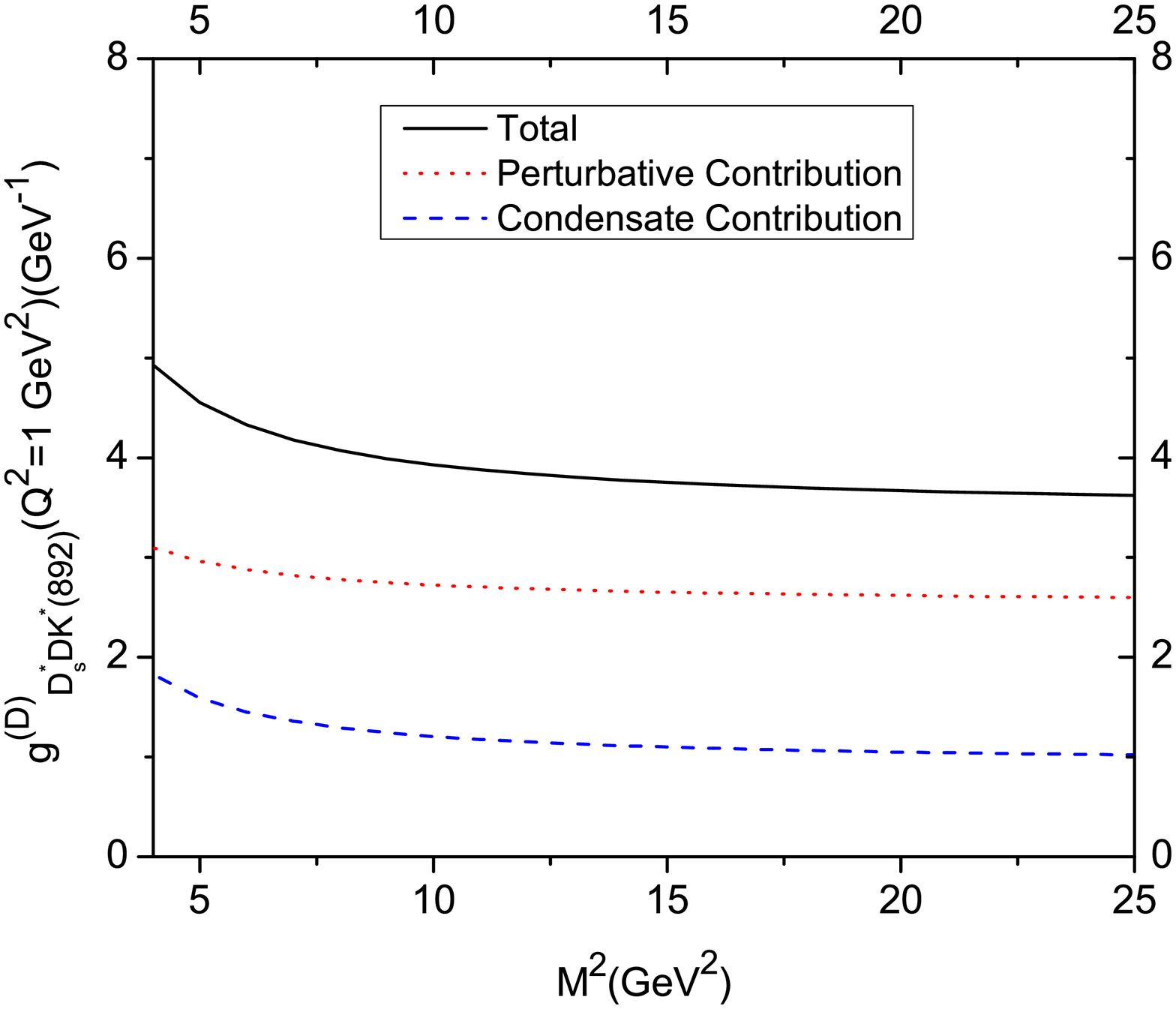}
\end{center}
\caption{$g^D_{D^{\ast}_sDK^{\ast}}(Q^2=1~GeV^2)$ as a function of
the Borel mass $M^2$. The continuum thresholds, $s_0=6.83~GeV^2$,
$s_0^{\prime}=2.54~GeV^2$ and ${M^{\prime}}^2=5~GeV^2$ have been
used. } \label{gDsDKsDoffMsq}
\end{figure}

\begin{figure}[h!]
\begin{center}
\includegraphics[width=13cm]{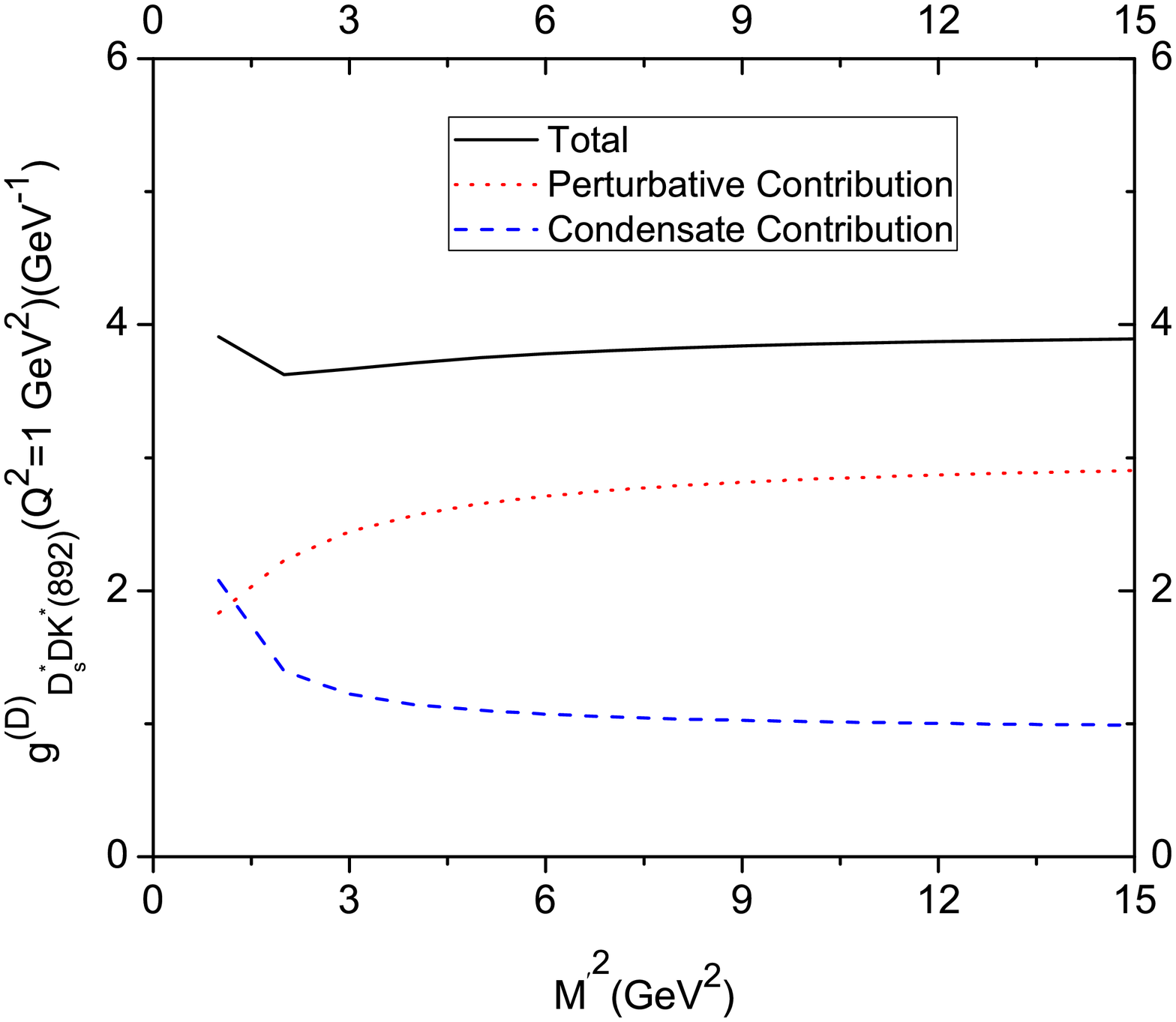}
\end{center}
\caption{$g^D_{D^{\ast}_sDK^{\ast}}(Q^2=1~GeV^2)$ as a function of
the Borel mass ${M^{\prime}}^2$. The continuum thresholds,
$s_0=6.83~GeV^2$, $s_0^{\prime}=2.54~GeV^2$ and ${M}^2=15~GeV^2$
have been used. } \label{gDsDKsDoffMpsq}
\end{figure}

\begin{figure}[h!]
\begin{center}
\includegraphics[width=13cm]{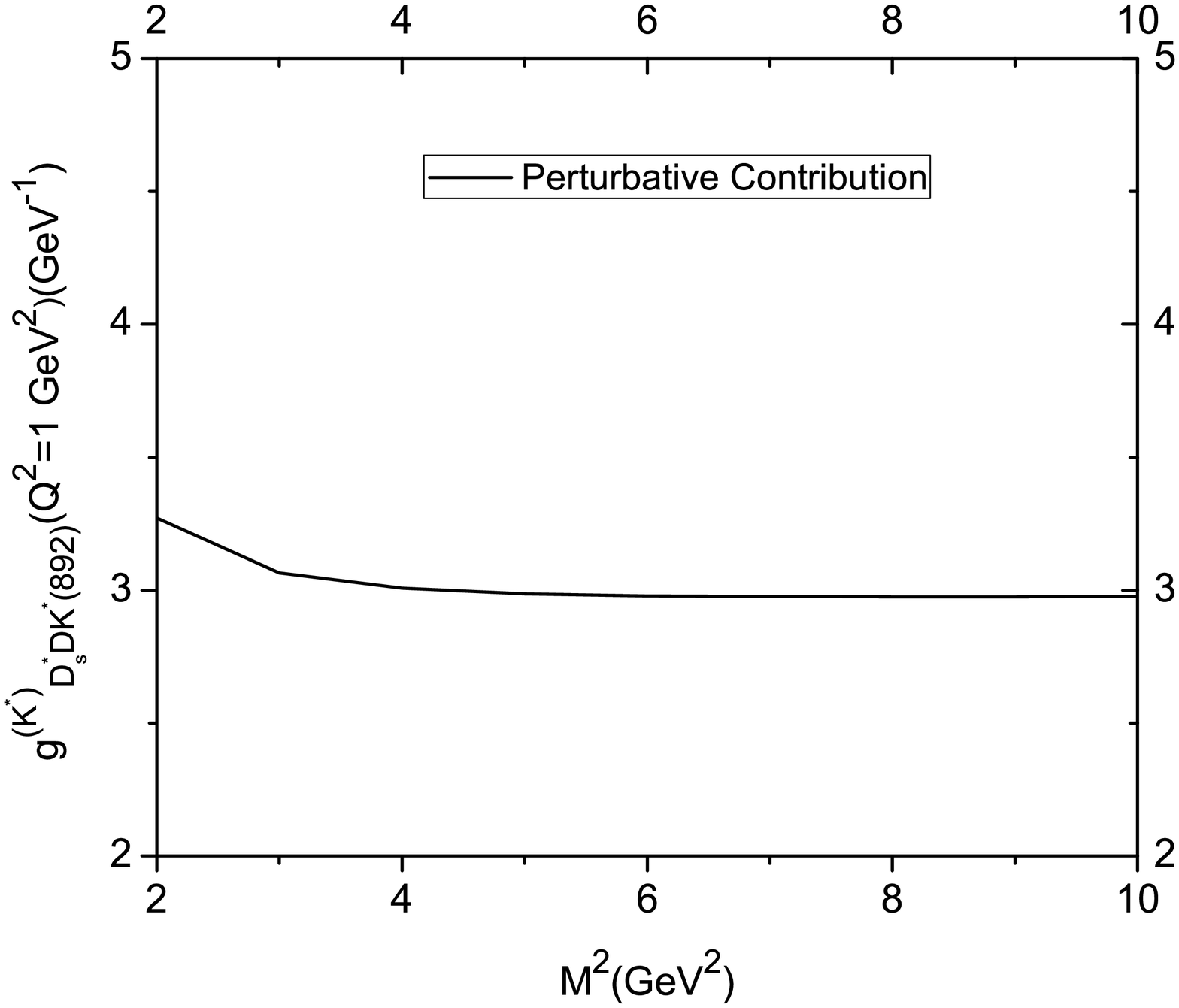}
\end{center}
\caption{$g^{K^{\ast}}_{D^{\ast}_sDK^{\ast}}(Q^2=1~GeV^2)$ as a
function of the Borel mass $M^2$. The continuum thresholds,
$s_0=6.83~GeV^2$, $s_0^{\prime}=6.57~GeV^2$ and
${M^{\prime}}^2=5~GeV^2$ have been used. } \label{gDsDKsKsoffMsq}
\end{figure}

\begin{figure}[h!]
\begin{center}
\includegraphics[width=13cm]{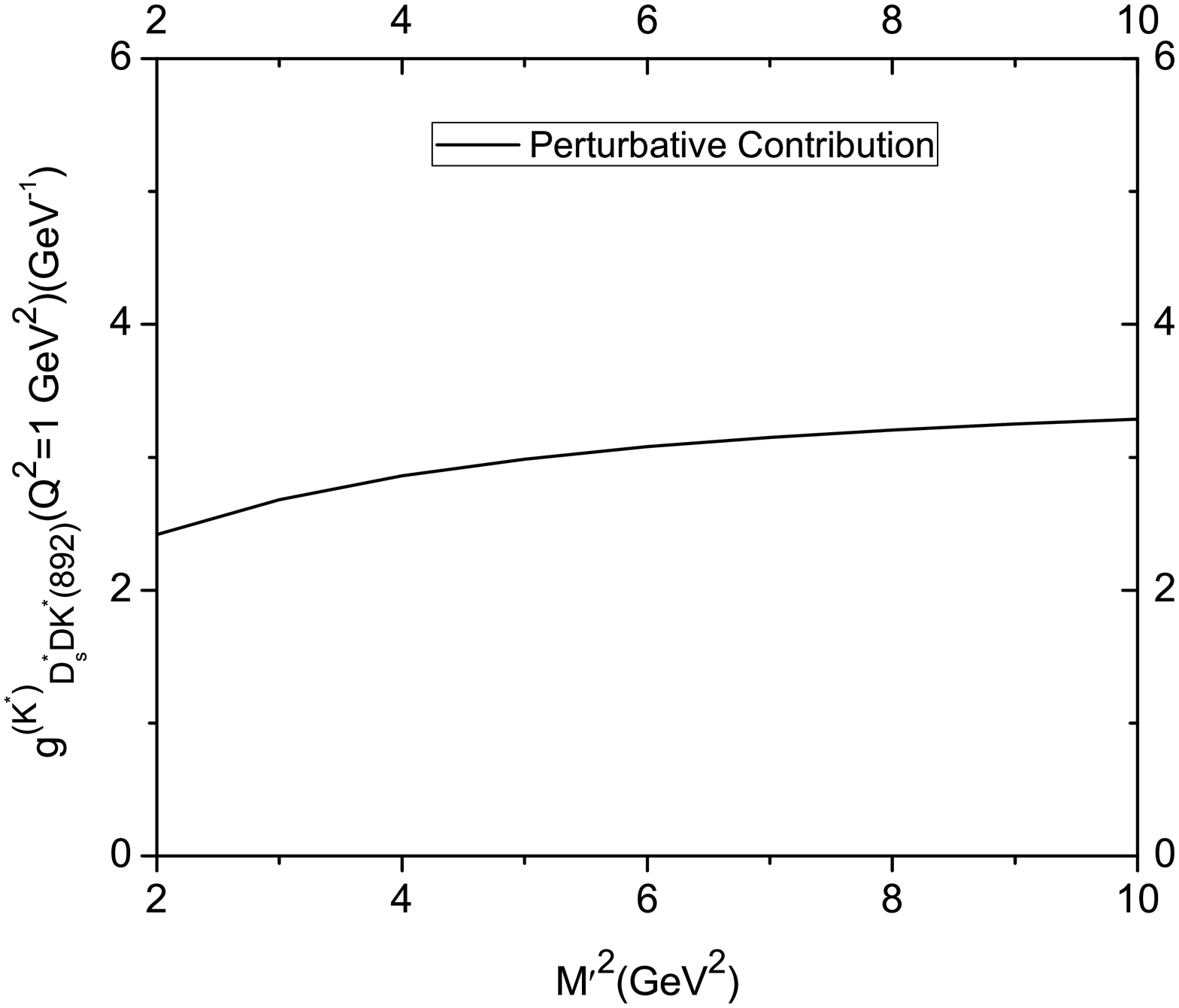}
\end{center}
\caption{$g^{K^{\ast}}_{D^{\ast}_sDK^{\ast}}(Q^2=1~GeV^2)$ as a
function of the Borel mass ${M^{\prime}}^2$. The continuum
thresholds, $s_0=6.83~GeV^2$, $s_0^{\prime}=6.57~GeV^2$ and
${M}^2=5~GeV^2$ have been used. } \label{gDsDKsKsoffMpsq}
\end{figure}

\begin{figure}[h!]
\begin{center}
\includegraphics[width=13cm]{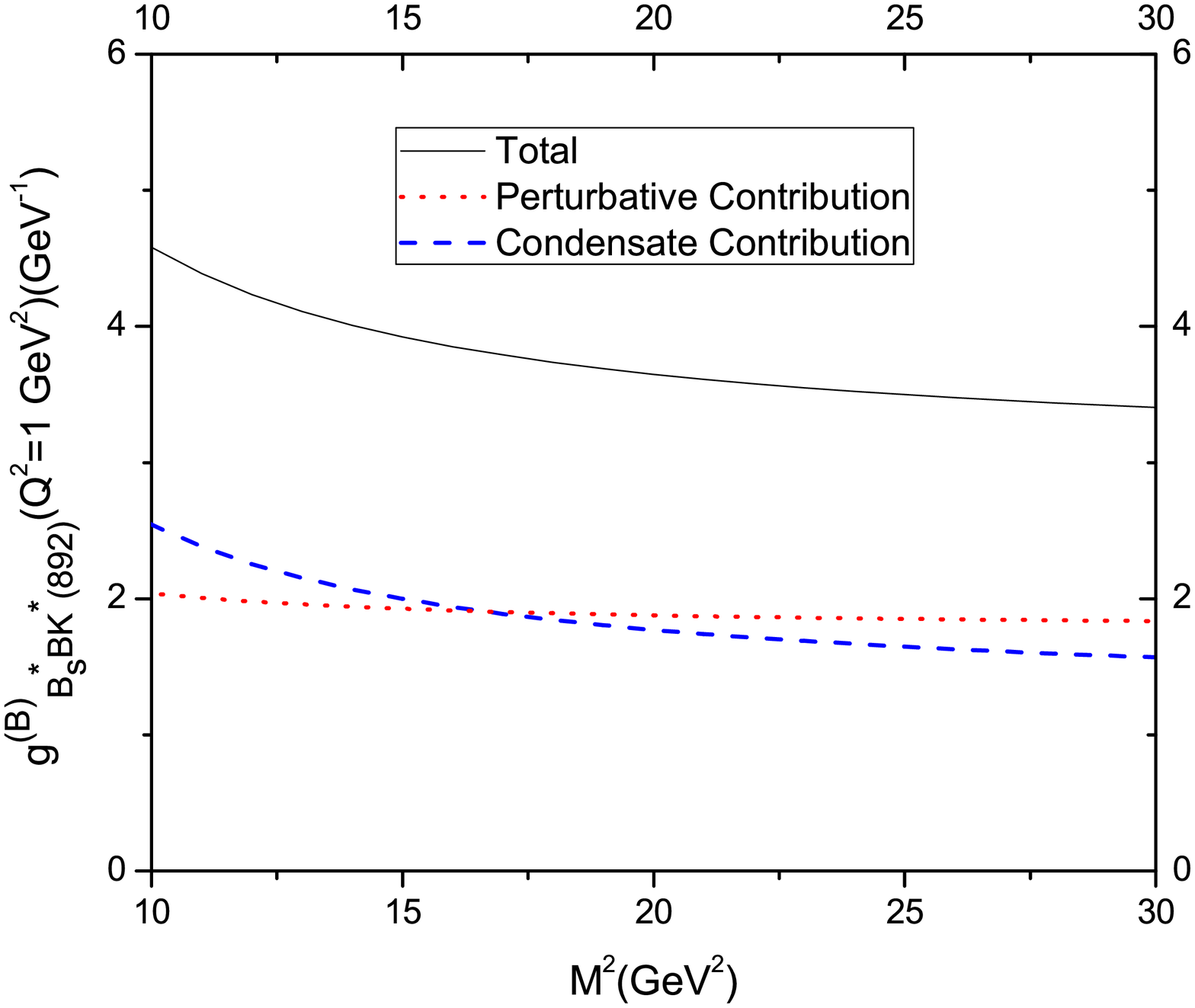}
\end{center}
\caption{$g^B_{B^{\ast}_sBK^{\ast}}(Q^2=1~GeV^2)$ as a function of
the Borel mass $M^2$. The continuum thresholds, $s_0=34.99~GeV^2$,
$s_0^{\prime}=2.54~GeV^2$ and ${M^{\prime}}^2=10~GeV^2$ have been
used. } \label{gBsBKsBoffMsq}
\end{figure}

 The sum rules for the strong coupling constants contain also four auxiliary
parameters, namely the Borel mass parameters, $M^2$ and ${M^{\prime}}^2$ and the continuum thresholds, $s_0$ and $s_0^{\prime}$. Since these
parameters are not physical quantities, our results should be
independent of them. Therefore, we look for working regions at which the dependence of coupling constants on these auxiliary parameters are weak. The working
 regions for the Borel mass parameters $M^2$ and
${M^{\prime}}^2$ are determined requiring that both the
contributions of the higher states and continuum are sufficiently
suppressed and the contributions coming from higher dimensions are small. As a result, we obtain, $8~GeV^2\leq M^2\leq25~GeV^2$ and
 $3~GeV^2\leq M'^2\leq15~GeV^2$ for $D$ off-shell, and $4~GeV^2\leq M^2\leq10~GeV^2$ and
 $3~GeV^2\leq M'^2\leq9~GeV^2$ for $K^{\ast}$ off-shell associated with the
$D_s^{\ast}D K^{\ast}(892)$ vertex. Similarly, the regions, $14~GeV^2\leq M^2\leq30~GeV^2$ and
 $5~GeV^2\leq M'^2\leq20~GeV^2$ for $B$ off-shell, and $5~GeV^2\leq M^2\leq20~GeV^2$ and
 $5~GeV^2\leq M'^2\leq15~GeV^2$ for $K^{\ast}$ off-shell are obtained for the $B_s^{\ast}B K^{\ast}(892)$ vertex.
The dependence of considered coupling constants on Borel parameters for different cases are
shown in Figs.(\ref{gDsDKsDoffMsq}-\ref{gDsDKsKsoffMpsq}) and (\ref{gBsBKsBoffMsq}-\ref{gBsBKsKsoffMsq}). From these figures, we see a good stability of the results
with respect to the Borel mass parameters in the working regions.
The continuum thresholds, $s_0$ and
$s_0^{\prime}$ are not completely arbitrary but they are correlated to the
energy of the first excited states with the same quantum numbers. Our numerical calculations lead to the following regions for the continuum thresholds in $s$ and $s'$ channels for different cases:
$(m_{D_s^{\ast}(B_s^{\ast}})+0.3)^2\leq s_0
\leq(m_{D_s^{\ast}(B_s^{\ast}})+0.5)^2$ in $s$ channel for both off-shell cases and two vertexes, and $(m_{D(B)}+0.3)^2\leq
s_0^{\prime}\leq (m_{D(B)}+0.7)^2$ and $(m_{K^{\ast}}+0.3)^2\leq
s_0^{\prime}\leq (m_{K^{\ast}}+0.7)^2$ for $K^{\ast}$ and $D(B)$ off-shell cases, respectively in $s'$ channel. Here, we should stress that the analysis of  sum rules in our 
work is based on, so called 
the standard procedure in QCD sum rules, i.e., the continuum thresholds are independent of Borel mass parameters and $q^2$. However, recently  it is believed that the standard procedure does not
render realistic errors and the continuum thresholds depend on Borel parameters and $q^2$ and this leads to some 
uncertainties (see for instance \cite{melikhov}). 

\begin{figure}[h!]
\begin{center}
\includegraphics[width=13cm]{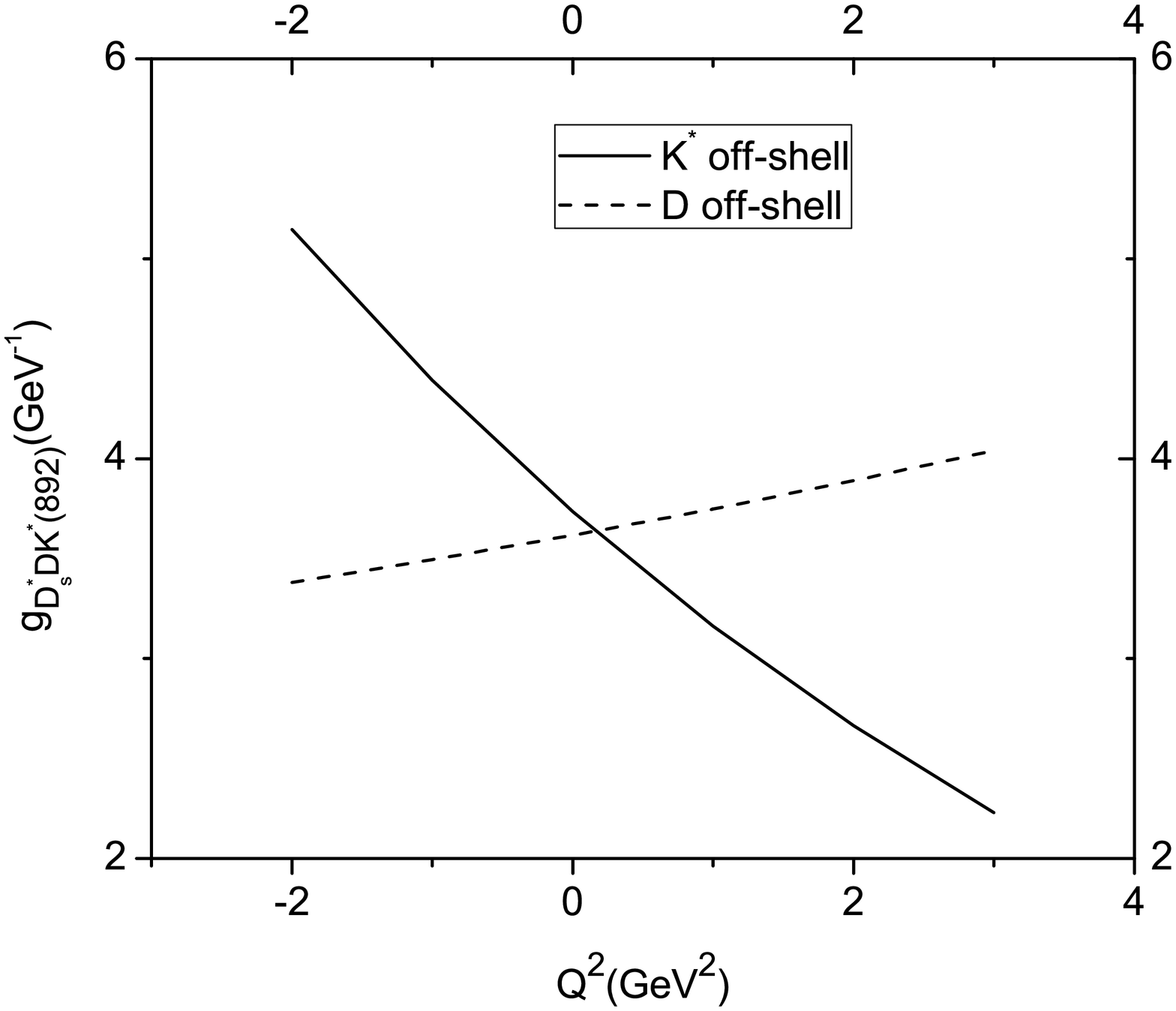}
\end{center}
\caption{$g_{D^{\ast}_sDK^{\ast}}$ as a function of $Q^2$.}
\label{gDsDKsQsq}
\end{figure}

Now, using the working region for auxiliary parameters and other input parameters, we would like to discuss the  behavior of the strong coupling constant form factors in terms of $q^2$.
In the case of  off-shell $D$ meson related to the $D^{\ast}_sDK^{\ast}$ vertex,
our numerical result  is described well by the following  mono-polar fit
 parametrization shown by the dashed line in
Fig. (\ref{gDsDKsQsq}):
\begin{eqnarray}\label{DsDKsFitD0offShell}
g^{(D)}_{D_s^{\ast}DK^{\ast}}(Q^2)=\frac{-103.34}{Q^2-28.57},
\end{eqnarray}
where $Q^2=-q^2$. The coupling constants are defined as the values
of the form factors at $Q^2=-m_{meson}^2$ (see also
\cite{BORodrigues}), where $m_{meson }$ is the mass of the on shell
meson. Using $Q^2=-m_D^2$ in Eq. (\ref{DsDKsFitD0offShell}), the
coupling constant for off-shell $D$ is obtained as:
$g^D_{D^{\ast}_sDK^{\ast}}=3.23~GeV^{-1}$. The result for an
off-shell $K^{\ast}$ meson can be well fitted by the exponential
parametrization presented by solid line in Fig. (\ref{gDsDKsQsq}) ,
\begin{eqnarray}\label{DsDKsFitKsoffShell}
g^{(K^{\ast})}_{D_s^{\ast}DK^{\ast}}(Q^2)=4.44~
e^{\frac{-Q^2}{7.24}}-0.70.
\end{eqnarray}
Using $Q^2=-m_{K^{\ast}}^2$ in Eq. (\ref{DsDKsFitKsoffShell}), the
$g^{K^{\ast}}_{D^{\ast}_sDK^{\ast}}=4.25~GeV^{-1}$ is obtained.
Taking the average of two above obtained values, finally we get the
value of the $g_{D^{\ast}DK^{\ast}}$ coupling constant as:
\begin{eqnarray}\label{CoupConstDsDKs}
g_{D^{\ast}_sDK^{\ast}}=(3.74\pm1.38)~GeV^{-1}.
\end{eqnarray}
From figure  (\ref{gDsDKsQsq}) it is also clear that the  form factor, $g^D_{D^{\ast}_sDK^{\ast}}$ is more stable comparing to $g^{K^{\ast}}_{D^{\ast}_sDK^{\ast}}$ with respect to the $Q^2$. The similar observation
 has also obtained in \cite{BORodrigues} in analysis of the $D^*D\rho$ vertex. In our case, the two form factors coincide at $Q^2=0.1612~GeV^2$ and have the value $3.64 ~GeV^{-1}$ very close to the
 value obtained taking average of the coupling constants for two off-shell cases at $Q^2=-m_{meson}^2$.

\begin{figure}[h!]
\begin{center}
\includegraphics[width=13cm]{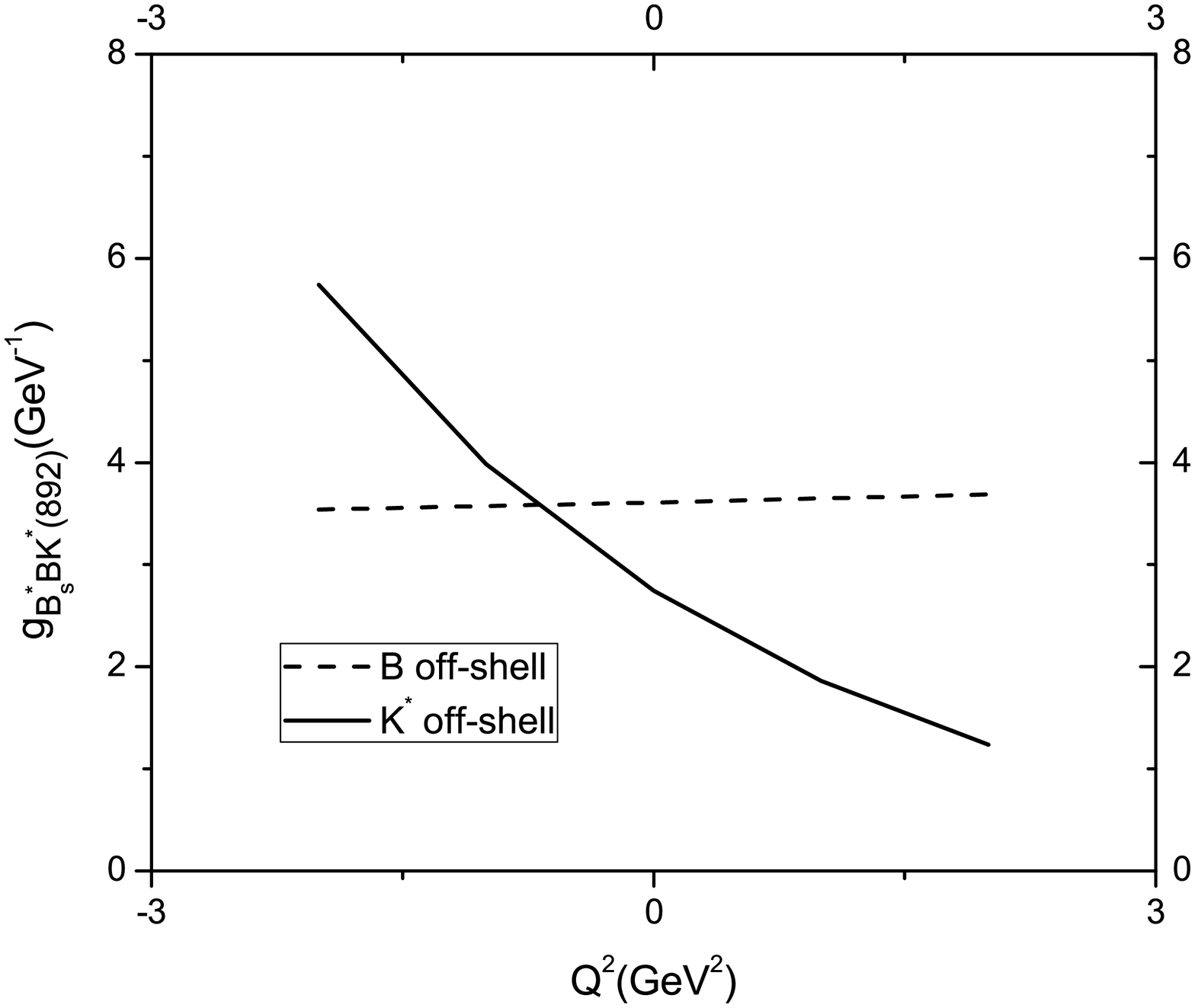}
\end{center}
\caption{$g_{B^{\ast}_sBK^{\ast}}$ as a function of $Q^2$.}
\label{gBsBKsQsq}
\end{figure}

Similarly, for $B^{\ast}_sBK^{\ast}$
vertex, our result for $B$ off-shell is better extrapolated by the mono-polar
 fit parametrization,
\begin{eqnarray}\label{BsBKsFitB0offShell}
g^{(B)}_{B_s^{\ast}BK^{\ast}}(Q^2)=\frac{-354.37}{Q^2-98.14},
\end{eqnarray}
presented by dashed line in Fig. (\ref{gBsBKsQsq}) and for $K^{\ast}$ off-shell case, the parametrization
\begin{eqnarray}\label{BsBKsFitKsoffShell}
g^{(K^{\ast})}_{B_s^{\ast}BK^{\ast}}(Q^2)=3.02~
e^{\frac{-Q^2}{2.90}}-0.28,
\end{eqnarray}
shown by the solid line in  Fig. (\ref{gBsBKsQsq}),  describes better the results in terms of $Q^2$.
 Using $Q^2=-m_B^2$ in Eq. (\ref{BsBKsFitB0offShell}),
the coupling constant is obtained as
$g^B_{B^{\ast}_sBK^{\ast}}=2.78~GeV^{-1}$. Also,
$g^{K^{\ast}}_{B^{\ast}_sBK^{\ast}}=3.69~GeV^{-1}$ is obtained at
$Q^2=-m_{K^{\ast}}^2$ in Eq. (\ref{BsBKsFitKsoffShell}). Taking the average of these results, we get,
\begin{eqnarray}\label{CoupConstDsDKs}
g_{B_s^{\ast}BK^{\ast}}=(3.24\pm1.08)~GeV^{-1}.
\end{eqnarray}
The errors in the results are due to the uncertainties in determination of the working regions for the auxiliary parameters as well as the errors in the input parameters. From the figure Fig. (\ref{gBsBKsQsq}), we
also deduce that the heavier is the off-shell meson, the more stable is its coupling
form factor in terms of $Q^2$. From this figure, we also see that the  two form factors related to the $B^{\ast}_sBK^{\ast}$ vertex coincide at $Q^2=-0.7152~GeV^2$ and have the value $3.58 ~GeV^{-1}$ also  close to the
 value obtained taking average of the corresponding coupling constants for two off-shell cases at $Q^2=-m_{meson}^2$.

\begin{figure}[h!]
\begin{center}
\includegraphics[width=13cm]{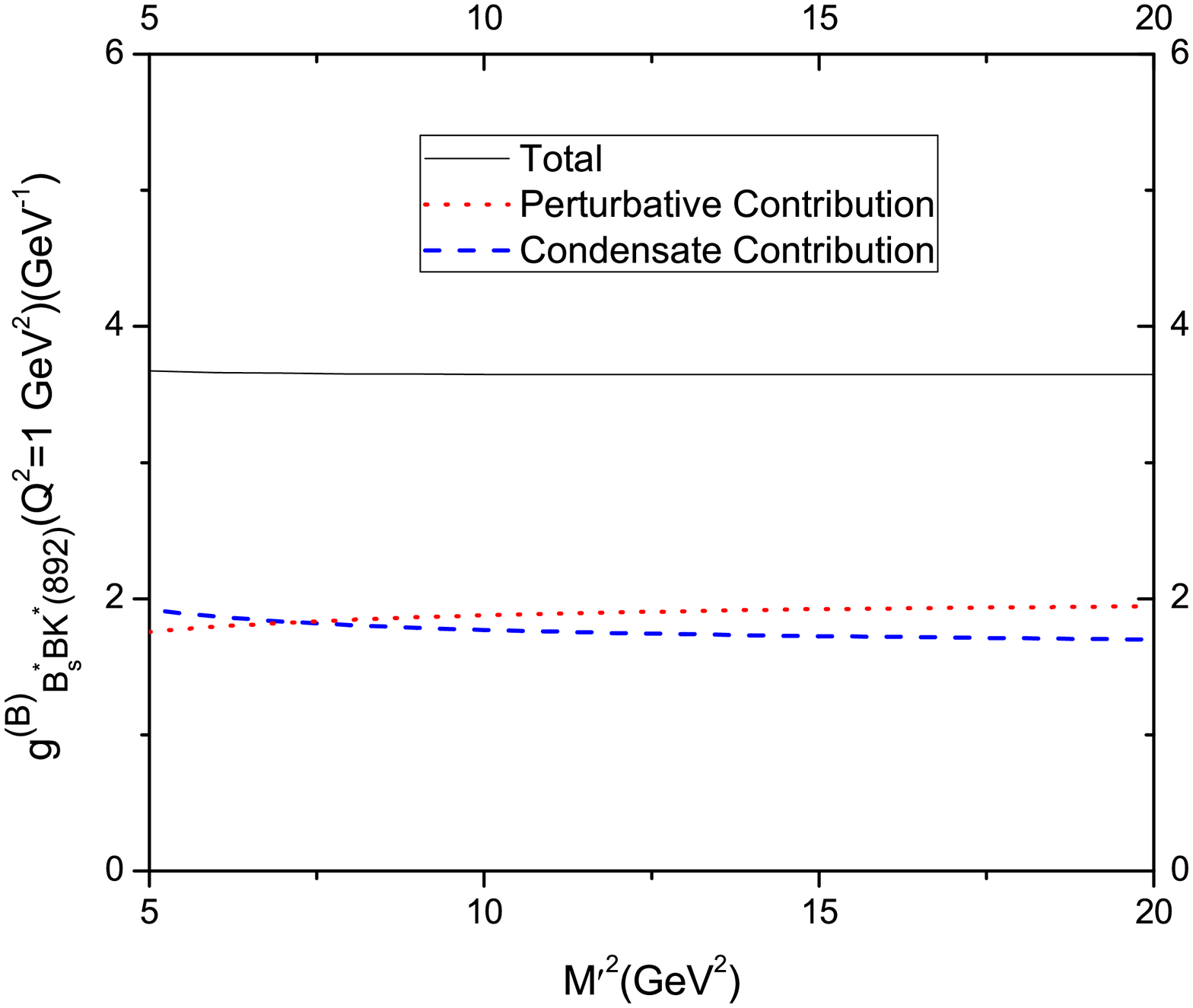}
\end{center}
\caption{$g^{K^{\ast}}_{B^{\ast}_sBK^{\ast}}(Q^2=1~GeV^2)$ as a
function of the Borel mass ${M^{\prime}}^2$. The continuum
thresholds, $s_0=34.99~GeV^2$, $s_0^{\prime}=2.54~GeV^2$ and
${M}^2=20~GeV^2$ have been used. } \label{gBsBKsBoffMpsq}
\end{figure}
\begin{figure}[h!]
\begin{center}
\includegraphics[width=13cm]{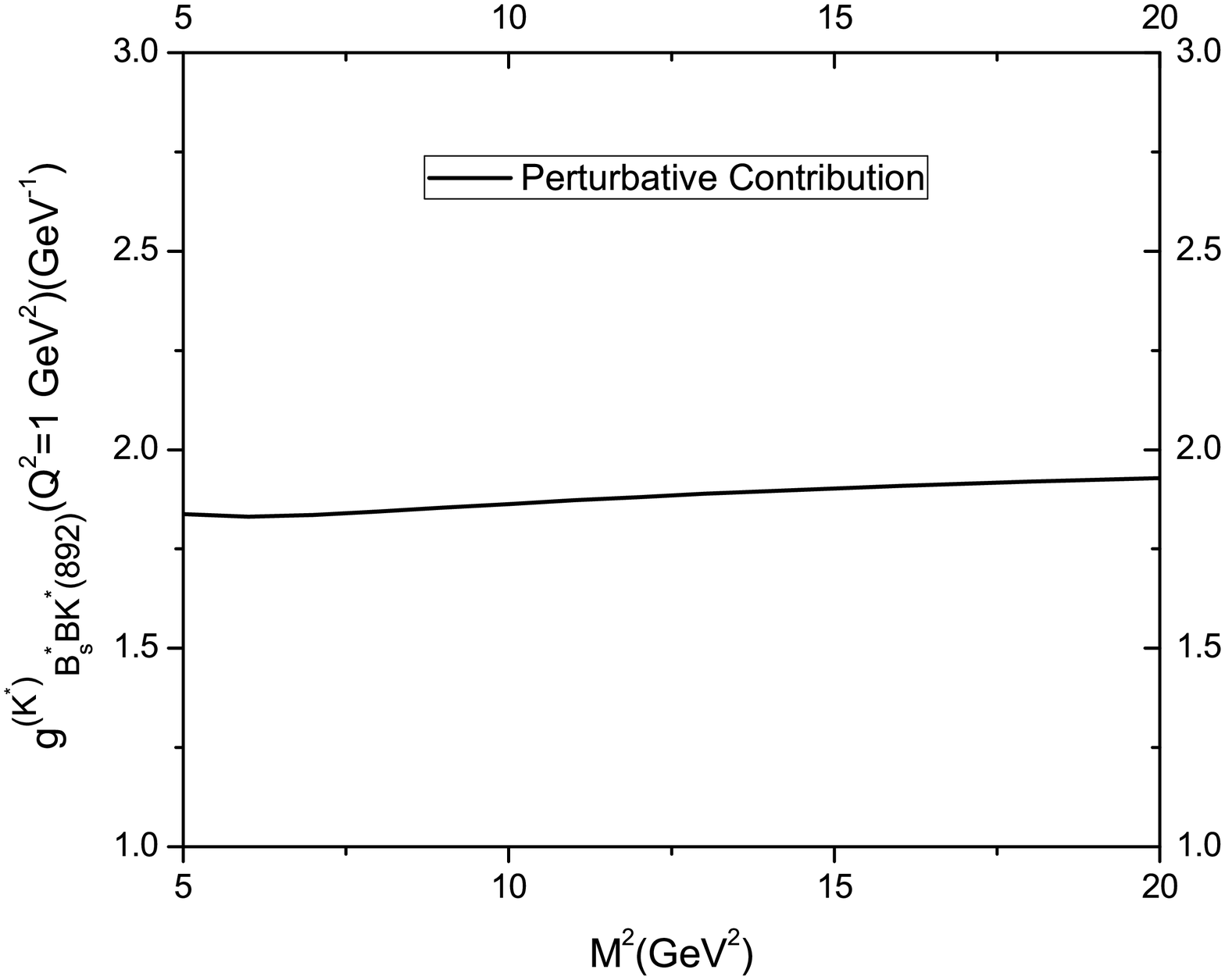}
\end{center}
\caption{$g^{K^{\ast}}_{B^{\ast}_sBK^{\ast}}(Q^2=1~GeV^2)$ as a
function of the Borel mass $M^2$. The continuum thresholds,
$s_0=34.99~GeV^2$, $s_0^{\prime}=35.75~GeV^2$ and
${M^{\prime}}^2=8~GeV^2$ have been used. } \label{gBsBKsKsoffMsq}
\end{figure}
\begin{figure}[h!]
\begin{center}
\includegraphics[width=13cm]{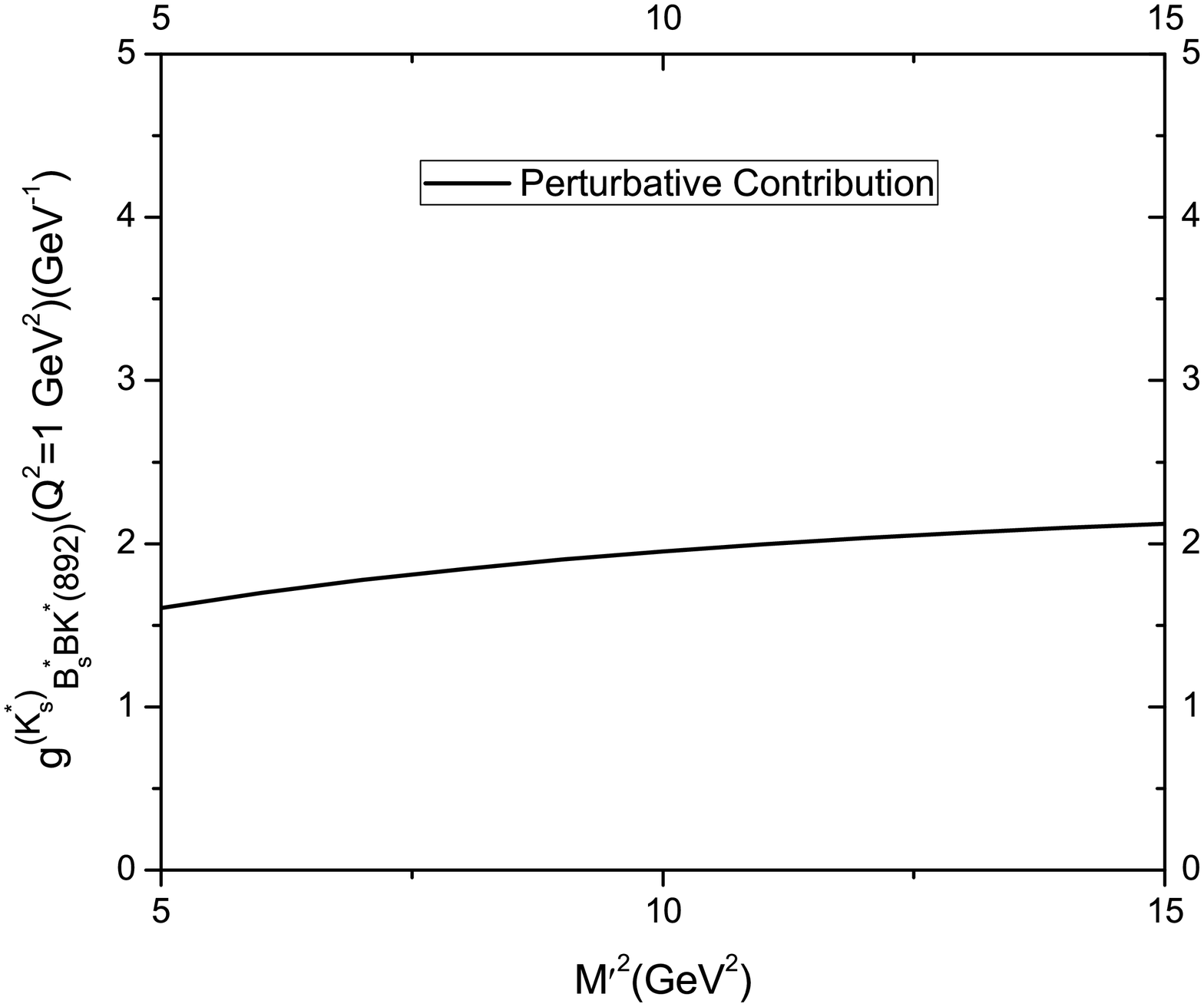}
\end{center}
\caption{$g^{K^{\ast}}_{B^{\ast}_sBK^{\ast}}(Q^2=1~GeV^2)$ as a
function of the Borel mass ${M^{\prime}}^2$. The continuum
thresholds, $s_0=34.99~GeV^2$, $s_0^{\prime}=35.75~GeV^2$ and
${M}^2=8~GeV^2$ have been used. } \label{gBsBKsKsoffMsq}
\end{figure}

\section{Acknowledgement}
This work has been  supported partly by the Scientific and Technological Research
Council of Turkey (TUBITAK) under research project No: 110T284.

\end{document}